\documentclass[prb,showpacs,twocolumn,superscriptaddress,aps,a4paper]{revtex4-1}
\usepackage{pstricks,pst-node,pst-text,pst-3d,graphpap,pst-plot}
\usepackage{dcolumn}
\usepackage{amsmath}
\usepackage{graphicx}
\usepackage{latexsym}
\usepackage{amsfonts}
\usepackage{amssymb}
\DeclareGraphicsExtensions{.pdf,.gif,.jpg}
\newcommand{\be}{\begin{equation}}
\newcommand{\ee}{\end{equation}}
\newcommand{\beq}{\begin{eqnarray}}
\newcommand{\eeq}{\end{eqnarray}}

\begin{document}

\def\bbe{\mbox{\boldmath $e$}}
\def\bbf{\mbox{\boldmath $f$}}
\def\bg{\mbox{\boldmath $g$}}
\def\bh{\mbox{\boldmath $h$}}
\def\bj{\mbox{\boldmath $j$}}
\def\bq{\mbox{\boldmath $q$}}
\def\bp{\mbox{\boldmath $p$}}
\def\br{\mbox{\boldmath $r$}}
\def\bz{\mbox{\boldmath $z$}}

\def\bfzero{\mbox{\boldmath $0$}}
\def\bfone{\mbox{\boldmath $1$}}

\def\dr{{\rm d}}

\def\tb{\bar{t}}
\def\zb{\bar{z}}

\def\tgb{\bar{\tau}}

\def\bC{\mbox{\boldmath $C$}}
\def\bG{\mbox{\boldmath $G$}}
\def\bH{\mbox{\boldmath $H$}}
\def\bK{\mbox{\boldmath $K$}}
\def\bM{\mbox{\boldmath $M$}}
\def\bN{\mbox{\boldmath $N$}}
\def\bO{\mbox{\boldmath $O$}}
\def\bQ{\mbox{\boldmath $Q$}}
\def\bR{\mbox{\boldmath $R$}}
\def\bS{\mbox{\boldmath $S$}}
\def\bT{\mbox{\boldmath $T$}}
\def\bU{\mbox{\boldmath $U$}}
\def\bV{\mbox{\boldmath $V$}}
\def\bZ{\mbox{\boldmath $Z$}}

\def\hH{\mbox{$\hat{H}$}}

\def\bcalS{\mbox{\boldmath $\mathcal{S}$}}
\def\bcalG{\mbox{\boldmath $\mathcal{G}$}}
\def\bcalE{\mbox{\boldmath $\mathcal{E}$}}

\def\bgG{\mbox{\boldmath $\Gamma$}}
\def\bgL{\mbox{\boldmath $\Lambda$}}
\def\bgS{\mbox{\boldmath $\Sigma$}}

\def\bgr{\mbox{\boldmath $\rho$}}
\def\bgs{\mbox{\boldmath $\sigma$}}

\def\a{\alpha}
\def\b{\beta}
\def\g{\gamma}
\def\G{\Gamma}
\def\d{\delta}
\def\D{\Delta}
\def\e{\epsilon}
\def\ve{\varepsilon}
\def\z{\zeta}
\def\h{\eta}
\def\th{\theta}
\def\k{\kappa}
\def\l{\lambda}
\def\L{\Lambda}
\def\m{\mu}
\def\n{\nu}
\def\x{\xi}
\def\X{\Xi}
\def\p{\pi}
\def\P{\Pi}
\def\r{\rho}
\def\s{\sigma}
\def\S{\Sigma}
\def\t{\tau}
\def\f{\phi}
\def\vf{\varphi}
\def\F{\Phi}
\def\c{\chi}
\def\w{\omega}
\def\W{\Omega}
\def\Q{\Psi}
\def\q{\psi}

\def\ua{\uparrow}
\def\da{\downarrow}
\def\de{\partial}
\def\inf{\infty}
\def\ra{\rightarrow}
\def\bra{\langle}
\def\ket{\rangle}
\def\grad{\mbox{\boldmath $\nabla$}}
\def\Tr{{\rm Tr}}
\def\hc{{\rm h.c.}}

\title{Magnetically induced pumping and memory storage in  quantum
rings}

\author{Michele Cini and Enrico Perfetto}
\affiliation{Dipartimento di Fisica, Universit\`a di Roma Tor
Vergata, Via della Ricerca Scientifica 1, 00133 Rome, Italy}
\affiliation{Istituto Nazionale
di Fisica Nucleare, Laboratori Nazionali di Frascati, Via E. Fermi
40, 00044 Frascati, Italy}


\begin{abstract}
Nanoscopic rings pierced by external magnetic fields and
asymmetrically connected to wires
behave in  sharp contrast with  classical 
expectations.
By studying the real-time evolution of  tight-binding models in
different geometries, we show that the
creation of  a magnetic dipole by a bias-induced current is a process
that can be
reversed: connected rings excited by an internal ac flux
produce ballistic currents in the external wires.
In particular we point out that, by employing suitable flux protocols,
single-parameter nonadiabatic pumping can be achieved, and an
arbitrary amount of
charge can be transferred from one side to the other.
We also propose a set up that could serve a memory device,
in which both the operations of {\it writing} and {\it erasing}
can be efficiently performed.

\end{abstract}


\maketitle
\date{\today}

\section{Introduction}

\label{secI}

The periodic modulation of the parameters of a mesoscopic system can
generate a finite dc current in absence of an
applied bias\cite{thouless}. This phenomenon is known as quantum pump
effect.
The quantum pumping has  attracted considerable attention during the
last decade
since it may provide a novel way  to
define a better current standard\cite{standard}, to produce
spin currents\cite{spin}, and  to realize
memory storage\cite{romeo}.
In a seminal work Brouwer\cite{brouwer} has shown that
in case of adiabatic modulations
at least two  parameters
(such as gate voltages) are needed to obtain a non-vanishing pumped
current. The connection of his arguments with the Berry phase is well
understood\cite{niu}.
An intriguing challenge in this field involves, however, the
incorporation
of non-adiabatic effects.
Simple schemes  for pumping
beyond the adiabatic approximation
in Floquet space were provided in Refs. \onlinecite{camalet,torres} . In
this space, the phase difference between
 two parameters was shown to provide a way to accumulate
a nonvanishing phase shift along a closed trajectory in a
way that is analogous to a magnetic flux. This motivated the proposal
of a quantum pump with a single time-dependent potential
and where the left-right symmetry is broken by a magnetic
field\cite{torres}.
In Ref. \onlinecite{stef} an alternative method based on real-time
simulations was proposed to study nonadiabatic effects
on the pumping properties of nanoscale junctions.

In the present study  we predict the
feasibility of single-parameter
non-adiabatic pumping in a circuit
containing a laterally connected ballistic ring pierced by
a time-dependent magnetic field\cite{lili}. 
We show that some aspects of this finding
are unexpected semiclassically  but can be rationalized in terms of a
recent theory\cite{iring} of current-induced magnetic moments
of quantum rings. As a further nonconventional application of the
same effect, we also
illustrate how the driven rings could be used
as memory devices, in which the characteristic charging and
discharging
timescales can be tailored in order to achieve efficient
writing/erasing protocols.

The paper is organized as follows.
In the next Section we introduce the tight binding model
describing the quantum ring connected to two metallic leads,
and elucidate the numerical method employed to
perform the real-time simulations.
In Section \ref{secIII} the numerical results
for the non-adiabatic pumping are presented.
We show two different setups to produce a pumped current
by varying the external magnetic field piercing the ring,
and we discuss the results in the light of a recent  theory
of quantum magnetic moments.
An interesting application beyond pumping
is proposed in Section \ref{secIV}, in which
it is shown that the
magnetically driven ring could also operate as a memory device.
Finally the summary and the main conclusions are drawn in
Section \ref{secV}.

\section{Model and numerical method}

\label{secII}

Using the tight-binding version of the partition-free approach to
quantum transport\cite{cini80} we
consider left ($L$) and right ($R$)
leads  connected to a
polygonal  ring with $N$ sites in presence of an ac magnetic fied.
The
time-dependent model Hamiltonian is\cite{units}:
\be
H(t)=H_{\rm ring}(t)+ H_{L}+H_{R} +H_{T}, \label{hamtot}
\ee
where the ring Hamiltonian reads
\be H_{\rm ring}(t)= \sum_{m,n=1}^{N}
h_{mn}(t)c^{\dag}_{m}c_{n} 
\ee
with hopping integrals
$h_{mn}=0$ if $m$ and $n$ are not nearest neighbors. The
time-dependent
magnetic field with flux $\phi(t)$ piercing the ring is
accounted for via the Peierls
prescription
$h_{m,n}(t)=t_{h}e^{\pm i \frac{2\pi \phi(t)}{N \phi_{0}}}\equiv
e^{\pm i
\frac{\alpha(t)}{N}} $ for nearest
neighbors, where  $\phi_{0}$ the flux quantum and the positive sign
holds
if the bond $m\rightarrow n$ runs clockwise and negative if
it is counterclockwise. This is  one of the many  ways
to insert a given flux in the ring:
less symmetric ways are not gauge-equivalent in the time-dependent
case but the
symmetric choice is the most natural.
The   leads are modeled by semi-infinite tight
binding chains described by the Hamiltonians $H_{L,R}$ with
nearest-neighbour hopping $t_{h}$ and on-site energies
$\epsilon_{L,R}$.
The ring is connected to the leads
via a tunneling Hamiltonian $H_{T}$ with hopping $t_{h}$ connecting
two nearest-neighbour sites of the ring denoted with $A$ and $B$ with
the ending sites of lead $L$ and $R$ respectively (see e.g. Fig.
\ref{fig2}).
At equilibrium the occupation of the system is determined by the chemical potential
$\mu$, which in the rest of the work is assumed to be zero.

 %

The numerical results below are obtained by computing
the exact time evolution of the system with a finite number of sites
$N_{\mathrm{lead}}$ in both $L$ and $R$ leads.  We first calculate the
equilibrium configuration by numerically diagonalizing
the $(N_{\mathrm{lead}}+N)\times (N_{\mathrm{lead}}+N)$ matrix
$H(0)$, and then we evolve the  equal-time lesser Green's function
of the system given by
\be
[G^{<}]_{nm}(t) \equiv G^{<}_{nm}(t)=i\bra c^{\dag}_{n}(t)
c_{m}(t)\ket.
\ee
In order to perform the time propagation we discretize the time
and calculate the matrix $G^{<}(t)$ according to
\begin{equation}
G^{<}(t_{n}) \approx   e^{-i H(t_{n})\Delta t} \,
G^{<}(t_{n-1}) \,  e^{i H(t_{n})\Delta t} \, ,
\end{equation}
where $t_{n}=n\Delta t$, $\Delta t$ is the time step, $n$ is a
positive integer and $G^{<}(0) = i f[H(0)]$, with $f$ the equilibrium
Fermi function.
The time-dependent current flowing between  sites $m$ and $n$ connected
by a bond with hopping integral
$h_{mn}$ is given by the average of the operator\cite{caroli}
\be
J_{mn}=-i( h_{mn}c^{\dag}_{m}c_{n}-h_{nm}c^{\dag}_{n}c_{m})
\label{scossa}
\ee
and is explicitly given by
\begin{equation}
J_{mn}(t)  = 2\mathrm{Re}
[ h_{mn} \, G^{<}_{nm}(t)] \, .
\label{current}
\end{equation}
Anagously the density at site $m$ reads
\be
n_{m}(t)= \mathrm{Re}[-iG^{<}_{mm}(t)]
\ee

This  approach allows us to reproduce the time
evolution of the infinite-lead system provided $N_{\mathrm{lead}}$
is large enough and the the propagation times do not exceed
the critical time $T_{\mathrm{lead}} \approx
N_{\mathrm{lead}}/t_{h}$\cite{perf}.

 %
 %


\section{Non-adiabatic pumping}

\label{secIII}

\begin{figure}[htbp]
\includegraphics*[width=.5\textwidth]{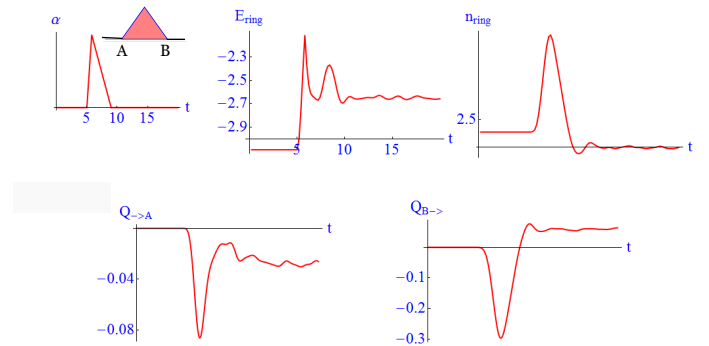}
\caption{Charging of a cluster with  $N=3$ sites at half filling,
connected to identical wires with   $\epsilon_{L}=\epsilon_{R}=0$ by
a burst of  flux. Top left:   time dependence
of the Peierls phase $\a(t)=2\pi \phi(t)/\phi_{0}$ inserted in the
ring. The shaded area denotes where the flux is inserted. The inset shows the geometry: wires are
joined to sites $A$ and $B$. Top center:  time dependence
of the expectation value of the ring Hamiltonian, showing that
the ring remains excited at the end. Top right:   total electron
number per spin in the ring, which remains charged at the end.
Bottom left:  electron
number (integrated current) from left wire to $A$. Bottom right:
electron
number from  $B$ to the right wire. Charge is expelled symmetrically
from the ring.
}
\label{fig2}
\end{figure}

First, we consider  the ring  connected to  identical leads  and
we seek a  method for pumping charge form lead $L$ to lead $R$ (or
viceversa) by employing the flux in the ring.
To make a pump, we need  a  cycle
whereby the system shifts some charge and  ends with no flux in the
ring, ready for starting again. If we insert some time-dependent
flux $\phi(t)$ and then reduce it to zero, this is equivalent to
using an electromotive force (EMF) first, e.g., clockwise and then
counterclockwise. In such numerical  experiments we did observe some
charge shift in the wires, although the total transmitted charge at
the end of the cycle
is close to zero,  irrespective of the detailed shape of $\phi(t)$.
In Figure \ref{fig2} we report the results of such a calculation for
the ransmitted charge
$Q_{\mathrm{B\to}}(t)=\int_{0}^{t}J_{B}(\tau)d\tau$ where $J_{B}$ is
the
current through $B$ to the right, and similarly
$Q_{\mathrm{\to A}}(t)=\int_{0}^{t}J_{A}(\tau)d\tau$.
We see that a positive number of electrons is going to the
right and a negative one enters from the left, and finally
the ring population is decreased without any significant
shift of charge from left to right.

\begin{figure}[htbp]
\includegraphics*[width=.5\textwidth]{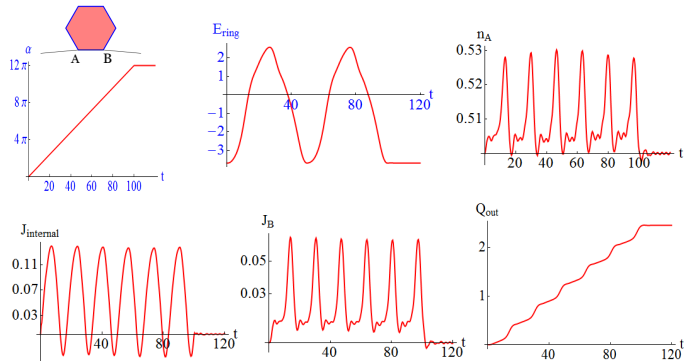}
\caption{(Color online) quantum pumping by a  6-atom
ring at half filling, connected as shown (wires are joined to
neighboring sites $A$ and $B$) to identical wires  and swallowing up
an integer number of fluxons. The shaded area denotes where the flux is inserted.
The time in abscissas is in
$\frac{\hbar}{t_{h}}$ units.
Top left:   the Peierls phase $\a(t)$ inserted in
the ring; the final value is 12$\pi$. Top center: the expectation
value of the ring Hamiltonian. Top right: occupation number at
site $A.$  Bottom left:
current $J_{\mathrm{internal}}$ flowing through all the internal bonds.
Bottom center: current at
A in left wire. Bottom right: \textit{staircase} shaped  increase of the electron
number $Q_{\mathrm{out}}=Q_{\to A}=-Q_{B \to}$ pumped
from left to right into wires.
 The hight of the \textit{staircase}
decreases if the switching time increases. In  other terms, the
process is observed to be non-adiabatic.
}
\label{fig3}
\end{figure}

Thus in order to produce a net pumped charge, an alternative
protocol is needed. One can start from zero flux
and insert an integer number of flux quanta. In this way, the final Hamiltonian is equivalent to the initial one.  As one
can see in Figure \ref{fig3} this produces the desired effect in a
ring with $N=6$, i.e. the same
amount of charge is transferred at any cycle.
To understand the physical origin of the charge transfer one can
think of the ring as a
renormalised $A-B$ bond in the circuit. By inserting the
time-dependent flux in the ring, one is producing a complex effective
$A-B$ bond which creates a phase difference between the $L$ and $R$
leads. In the time-dependent case the phase difference entails an
effective level difference, equal to $\dot{\alpha},$ which is
the reason why a charge transfer takes place. This effect does not
decrease
with the number of sides $N$. Actually,
by repeating the
calculation with the same parameters but $N=18$ (not shown), one
obtains a very similar behavior with somewhat higher steps
(the staircase halts at about 3.0).  However, it  decreases
steadily if one increases the switch-on time of the flux
$T_{\mathrm{sw}}$.
Numerically we find that the height  of the staircase in Fig.2 goes
like $T_{\mathrm{sw}}^{-0.3}.$
Therefore the charge transfer is a non-adiabatic process.

Strictly speaking, this is not yet pumping since the flux changes linearly in time and this is equivalent to a constant electromotive  force inside the ring, rather than an AC one.  A strict example of pumping is presented below.

\begin{figure}[htbp]
\includegraphics*[width=.4\textwidth]{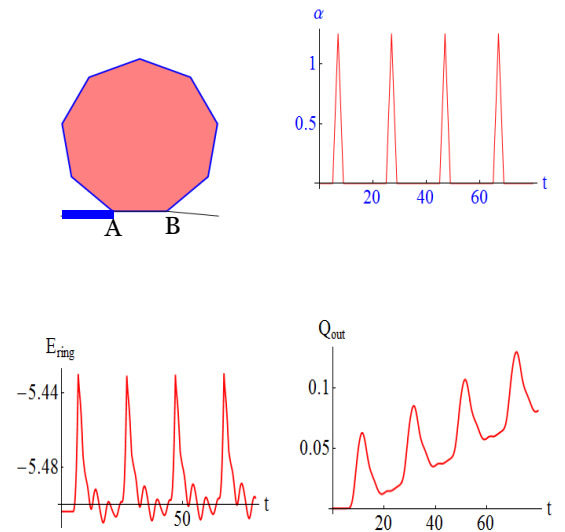}
\caption{(Color online) Pumping by  a  tight-binding  ring with
$N=9$
at half filling, connected to a junction with $\epsilon_{L}=2$,
$\epsilon_{R}=0$.
Top left:  sketch of the geometry (wires are joined to neighboring
sites $A$ and $B$ and the shaded area denotes where the flux is inserted.).
Top right:  time dependence
of the Peierls phase $\a(t)$.   Bottom left:
expectation value of the ring Hamiltonian  versus time.
Bottom right:
increase of the electron
number $Q_{\mathrm{out}}=Q_{\to A}=-Q_{B \to}$ pumped
from left to right into wires.}
\label{fig4}
\end{figure}

We considered rings connected to wires having different equilibrium
occupancy.
 If the atoms in the left wire have energy levels
$\epsilon_{L}=2$ while in the rest of the system the levels
are at $\epsilon=0$, one
can expect that when the EMF tends to pump charge
into the wires this will happen more effectively towards the right,
because of the energy barrier $\epsilon_L-\epsilon_R$  at the junction.
The phase is inserted as a series of
triangular pulses, each involving much less than $\phi_{0}$.
As one can see in Figure \ref{fig4},
this choice of the model parameters again produces the desired effect
for a ring with $N=9$.
The quasi-periodic behavior of the expectation value of the ring
Hamiltonian suggests that after each cycle the ring returns to
the same state.
The relaxation is actually incomplete, but the oscillations of
$E_{\mathrm{ring}}(t)$ between successive flux pulses around the
ground-state value are very small.
However at each cycle a net charge is thrown into the
external ballistic circuit.
The charge injection occurs mainly during the duration of the
flux pulses, and the  transient effects due to incomplete
relaxation do not reduce the pumping.

Symmetrically connected rings cannot have a magnetic moment, since  by time reversal one changes the signs of currents and magnetic field and obtains an equivalent problem. We are considering laterally connected rings because they are maximally asymmetric.  
We can  understand qualitatively how  the
device works. 
In the pumping
process,  an applied magnetic field threading the ring
excites a current in the external circuit. This  is the inverse
of the generation of a magnetic moment by the current excited by an
external bias.  The
two processes must be related, and this analogy 
led us to the present work.
In a recent\cite{iring} work we showed that the magnetic moment
generated by a bias-induced current is at least quadratic in the bias.
In the present time-dependent problem where
the roles of cause and effect are interchanged, we find that
the pumped charge is quadratic in the
magnetic flux. Our numerical calculations show that for any number $N$ of
sides the  charge $Q_{\mathrm{out}}$  pumped
at each cycle grows with the square of the
height $\alpha_{\mathrm{max}}$ of the triangular pulses in Figure
\ref{fig4}.
We notice that this is typical behavior of one parameter pumping, as
already found by Foa Torres \cite{torres} who considered an open dot driven
by  time-dependent gate voltages. If the response is linear
the Brouwer  theorem\cite{brouwer} implies adiabatic behavior and
only multi-parametric pumping can be obtained\cite{cohen}.   The absence of a linear contribution of the ring
magnetic moment in the applied bias\cite{iring} and
the realization of one-parameter non-adiabatic pumping are both
manifestations of the same quantum effect.

\begin{figure}[htbp]
\includegraphics*[width=.4\textwidth]{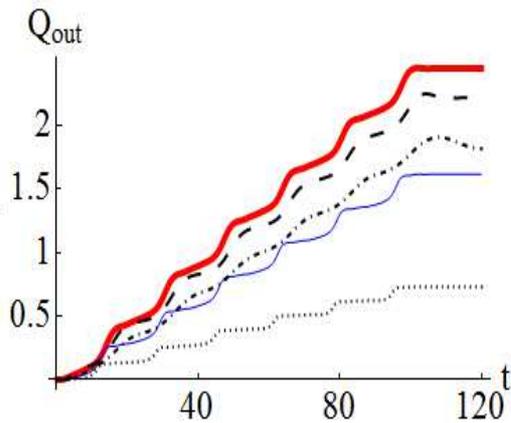}
\caption{(Color online)
 $Q_{\mathrm{out}}=Q_{\to A}=-Q_{B \to}$ versus time for the same geometry and flux
as in Figure  2 but different hopping in the ring. Thick line:
$t_{ring}=t_h$ as in Figure 2. Dashed line: $t_{ring}=t_{h}/2$.
Dash-dotted line: $t_{ring}=t_{h}/4$. Thin line:  $t_{ring}=2
t_h$.  Dotted line: $t_{ring}=4 t_h$.
 }
\label{fignuova}
\end{figure}


We have also studied the robustness of the above effect by varying
the model parameters. As one can see in Figure 4,  the
pumping  efficiency of the ring is optimal for  $t_{\mathrm{ring}}\sim t_h$.
An increase in $t_{\mathrm{ring}}$  implies that the electron prefers to
move around in the ring and this  causes  a reduced tranferred charge.
However the pumping also
decreases when the hopping $t_{\mathrm{ring}}$ is reduced, because the
circulating current is proportional to $t_{\mathrm{ring}}$.  This is
expected  since at the limit   $t_{\mathrm{ring}}=0$ the circuit is
interrupted, but the decrease is seen to be  rather slow and  for
$t_{\mathrm{ring}}=t_{h}/4$  the pumping is still comparable to the
$t_{\mathrm{ring}}=t_h$  case.  Overall we may conclude that the proposed pumping
is a robust effect, which survives to variations or
renormalizations of the parameters such as could result from
interactions in conductors which  are well described by a Fermi
liquid picture.

\begin{figure}[htbp]
\includegraphics*[width=.45\textwidth]{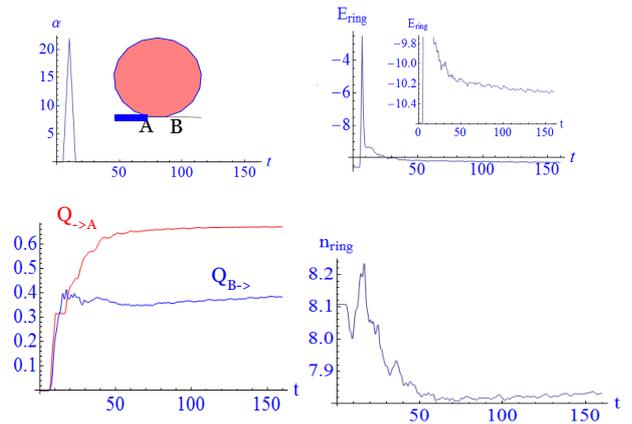}
\caption{(Color online) Effects of a strong magnetic
pulse on a ring with $N=17$ connected to a junction
($\epsilon_{L}=2$ while $\epsilon_{R}=0$). Top left: the geometry.
Top center: time dependence
of the  phase pulse $\a(t)$ inserted in the shaded area. Top right: expectation value of
the ring Hamiltonian. Bottom left: charge from $A$ to the left wire.
Bottom center: charge from the right wire to $B$. Bottom right:
electron population of the  ring, which remains charged, much more
than in Figure 1, thus keeping a memory of the pulse.}
\label{fig5}
\end{figure}

\begin{figure}[htbp]
\includegraphics*[width=.4\textwidth]{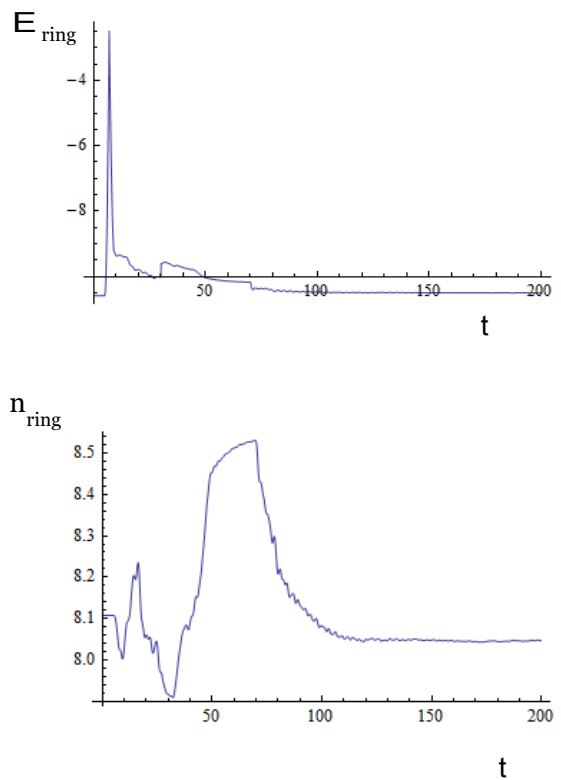}
\caption{(Color online) Same calculation as in Fig.\ref{fig5}
performed in
the 17-sided ring, but with the $A-B$ bond
cut between times $t=30$ and $t=70$. The ring energy and
occupation tend to return to the values they had at the beginning,
and the memory of the flux is thereby erased.
}
\label{fig6}
\end{figure}

\section{Memory storage}

\label{secIV}

The possibility of pumping a desired charge form one lead to the other
via a suitable protocol suggests that quantum rings can also be
used as memory devices.
In
Figure \ref{fig5} we report the outcomes of the insertion and removal
of  3.5 flux quanta into a  ring with $N=17$.
The on-site energies on the leads are $\epsilon_{L}=2$ while
$\epsilon_{R}=0$.  The disturbance
lasts $10/t_{h}$ time units. After the end of
the pulse the ring remains in an excited state, as one can
see by the expectation value of the ring Hamiltonian, which
does not return to the ground state value,
thus indicating that that the ring remains charged. According to the
above
discussion, when the ring is connected to an asymmetric junction, a
current
is generated in the wires.  From the bottom panels of Figure
\ref{fig5} one can see that the charge which is expelled to the $L$
wire is not totally compensated by the charge coming from the
$R$ wire. If the ring eventually returns to the original
charge state, it does so quite slowly compared to the time it takes to
get charged.  So we may say that the ring keeps memory of the pulse.
However if a memory is to be useful we must know a fast
mechanism to erase it. We have observed that sending a negative
pulse does not achieve that; rather, the effects tend to sum.
This is not surprising, since the charge shift is a quadratic
function
of the flux. On the other hand we have observed
flushing the ring with an external bias does not efficiently reduce
the charge on it nor does it bring the ring any closer to the ground
state energy. We found an efficient alternative, however, which
consists
in removing the $A-B$ bond for some tens of time units. The effects of
doing so are shown in Figure \ref{fig6}: the system becomes a mere
junction,
so the perturbation tends to delocalize. In this way the initial
condition is re-enabled, as required for a memory storage device.

\section{Conclusions}

\label{secV}

In conclusion we have invstigated the possibility of single-parameter
nonadiabatic pumping in a quantum circuit hosting a ballistic ring
pierced by a time-varyong magnetic field.
We adopted a time-dependent approach in which the electron dynamics
is calculated in real time.
We provide evidence that a pumped current can be generated in
asymmetric geometries, as well as if
integer number of flux quanta are inserted in time via a ramp
protocol.
Finally we have shown that driven quantum rings can also be used as
memory devices with the possibility of performing basic
writing/erasing operations. Our results extend those of a recently
proposed quantum
theory of magnetic moments in quantum rings\cite{iring}, and allow us
to conclude
that the absence of a ring
magnetic moment to first-order in the applied bias and
the existence of one-parameter non-adiabatic pumping are both
manifestations of the same quantum effect.


\begin{thebibliography}{10}
\bibitem{units}
In this work we will use atomic units and spin
indices will be omitted.
\end{thebibliography}
\end{document}